# Self-Assembled Room Temperature Multiferroic BiFeO$_3$-LiFe$_5$O$_8$ Nanocomposites

*Yogesh Sharma[†*], Radhe Agarwal[‡], Liam Collins[§], Qiang Zheng[†,$], Anton V. Ievlev[§], Raphael P. Hermann[†], Valentino R. Cooper[†], Santosh KC[†], Ilia N. Ivanov[§], Ram S. Katiyar[‡], Sergei V. Kalinin[§], Ho Nyung Lee[†], Seungbum Hong[#], Thomas Z. Ward[†*]*

[†]*Materials Science and Technology Division, Oak Ridge National Laboratory, Oak Ridge, TN 37831, USA*

[‡]*Department of Physics, University of Puerto Rico, San Juan, PR 00931, USA*

[§]*Center for Nanophase Materials Sciences, Oak Ridge National Laboratory, Oak Ridge, TN 37831, USA*

[$]*Department of Materials Science and Engineering, University of Tennessee, Knoxville, TN 37996, USA*

[#]*Department of Materials Science and Engineering, KAIST, Daejeon 34141, Republic of Korea*

*[sharmay@ornl.gov](mailto:sharmay@ornl.gov) and [wardtz@ornl.gov](mailto:wardtz@ornl.gov)*

ABSTRACT

Multiferroic materials have driven significant research interest due to their promising technological potential. Developing new room-temperature multiferroics and understanding their fundamental properties are important to reveal unanticipated physical phenomena and potential applications. Here, a new room temperature multiferroic nanocomposite comprised of an ordered ferrimagnetic spinel α-LiFe$_5$O$_8$ (LFO) and a ferroelectric perovskite BiFeO$_3$ (BFO) is presented. We observed that lithium (Li)-doping in BFO favors the formation of LFO spinel as a secondary phase during the synthesis of Li$_x$Bi$_{1-x}$FeO$_3$ nanoceramics. Multimodal functional and chemical imaging methods are used to map the relationship between doping-induced phase separation and local ferroic properties in both the BFO-LFO composite ceramics and self-assembled nanocomposite thin films. The energetics of phase separation in Li doped BFO and the formation of BFO-LFO composites is supported by first principles calculations. These findings shed light on Li's role in the formation of a functionally important room temperature multiferroic and open a new approach in the synthesis of light element doped nanocomposites.

KEYWORDS: Multiferroics, self-assembled nanocomposites, thin film nanostructures, lithium doping, nanoferroic properties, scanning probe microscopy.

Multiferroics have great potential in developing new low energy and low cost applications.[1–3] Particular interest has been given to materials systems where ferroelectricity and magnetism (ferro-, ferri-, and antiferro-magnetic orders) are both present.[4] These multiferroic materials can potentially host magnetoelectric effects, where the polarization P and the magnetization M respond to applied magnetic and electric fields.[4–7] The scarcity of single-phase room temperature multiferroic materials has led to the development of other synthesis approaches where different ferroic materials are combined through heterostructuring or by nanocompositing.[8–13] Synthesizing composite structures has enabled an attractive approach to designing new multiferroic materials.[14–18] In composite multiferroics, parent materials with different ferroic properties are combined.[15] In such nanocomposites, immiscibility gaps between oxide materials leads to phase-decomposition-based self-assembly of ferroelectric and magnetic phases.[14] Well-known examples of self-assembled multiferroic composites often consist of ferrimagnetic spinel-$CoFe_2O_4$ nanodomains embedded in a ferroelectric perovskite matrix, where the coupling between magnetization and polarization is mediated by the elastic strain between the two phases[14,19,20]. Self-assembled multiferroic nanocomposite functionality is often dominated by unique interface-mediated couplings that drive macroscopic properties.[21–28] Thus, designing new multiferroic nanocomposites requires an understanding of how domain structure and interaction dictate function, while synthesis requires the ability to control the relative phase compositions and morphologies.[15,17,21,24,25,29–35]

Light element Li doping is known to improve the piezoelectric properties in ferroelectric ceramics by shifting the morphotropic phase boundary to room temperature, however the mechanism driving this response is elusive.[36,37] Recently, Li doped bismuth ferrite ($BiFeO_3$) was reported to show room temperature ferromagnetism and spintronic functionality.[38,39] Again, the mechanism is unclear with questions remaining as to the preferred position of the Li atoms in the $BiFeO_3$ matrix (i.e., Bi or Fe substitution) and what role secondary nanoscale phases might play. In this work, Li doping is shown to control formation of magnetic and ferroelectric phase composition during the synthesis of $Li_xBi_{1-x}FeO_3$ ($x$ = 0, 0.03, 0.09) bulk ceramics. This new multiferroic nanocomposite consists of a perovskite BiFeO$_3$ matrix which hosts nanoscopic secondary phase inclusions of spinel LiFe$_5$O$_8$, with the spinel phase being solely responsible for the room temperature magnetic behavior. It is also demonstrated that these ceramic pellets can be used as pulsed laser deposition targets to allow synthesis of self-assembled BiFeO$_3$-LiFe$_5$O$_8$ thin film nanostructures where LiFe$_5$O$_8$ nanopillars are heteroepitaxially embedded in a single crystal BiFeO$_3$ matrix.

**RESULTS AND DISCUSSION**

Phase pure BiFeO$_3$ (BFO) and Li$_x$Bi$_{1-x}$FeO$_3$ [x = 0.03 (3Li-BFO) and x = 0.09 (9Li-BFO)] bulk ceramics are synthesized using conventional solid-state reaction (see methods). The 9Li-BFO sample is chosen for detailed experimental studies as the optimum spinel-perovskite concentration can be observed at x = 0.09 (see supporting information). Figure 1a shows the Rietveld refinement for the powder X-ray diffraction (XRD) pattern of Li$_{0.09}$Bi$_{0.91}$FeO$_3$ (9Li-BFO) sample. Besides the main BiFeO$_3$ phase with a rhombohedral structure (space group *R*3*c*, JCPDS file no. 71-2494), two other phases, *i.e.* the ordered spinel α-LiFe$_5$O$_8$ phase (space group *P*4$_3$32, JCPDS file no. 74-1726) and the Bi$_{12.5}$Fe$_{0.5}$O$_{20}$ phase (space group *I*23, JCPDS file no. 78-1543), exist in the sample. The low residual fitting values ($R_P$=0.062, $wR_P$=0.079) indicate high quality of full-pattern refinement with volume contents of the three phases estimated to be 78.8(5)% BiFeO$_3$ (BFO), 15.0(5)% LiFe$_5$O$_8$ (LFO), and 6.2(2)% Bi$_{12.5}$Fe$_{0.5}$O$_{20}$ (see supporting information for refinement details). The micro-Raman spectroscopy and mapping measurements were employed to further confirm the existence of a phase mixture and the distribution of these phases at the microscopic scale. Figure 1b shows a 15×7 μm$^2$ area on the surface of a 9Li-BFO ceramic pellet that was mapped by Raman spectroscopy across a grid having a 500 nm step size. The Raman mapped area is characterized by three types of spectra relating to BFO, Bi-excess BFO, and LFO phases (Figure 1b-c). It is worth noting that the observed phonon modes of the LFO phase perfectly match the expected Raman spectrum of bulk single crystal LFO.[40] The sharp, high intensity $A1$ mode at 124 cm$^{-1}$ indicates a nearly uniform breathing of the FeO$_4$ tetrahedra that is the result of an ordered $\alpha$ phase in the LFO spectrum.[40] The Raman spectrum of Bi-excess BFO differs in the relative line intensities compared to the rhombohedral BFO. However, the observation of additional modes at 358, 490, and 530 cm$^{-1}$ of broader linewidth suggests the presence of Bi$_{12.5}$Fe$_{0.5}$O$_{20}$ phase.[41] The Raman results strongly agree with the X-ray findings that the BFO-LFO composites, together with the small fraction of Bi$_{12.5}$Fe$_{0.5}$O$_{20}$ phase, can be revealed at room temperature in the Li doped BFO ceramics.

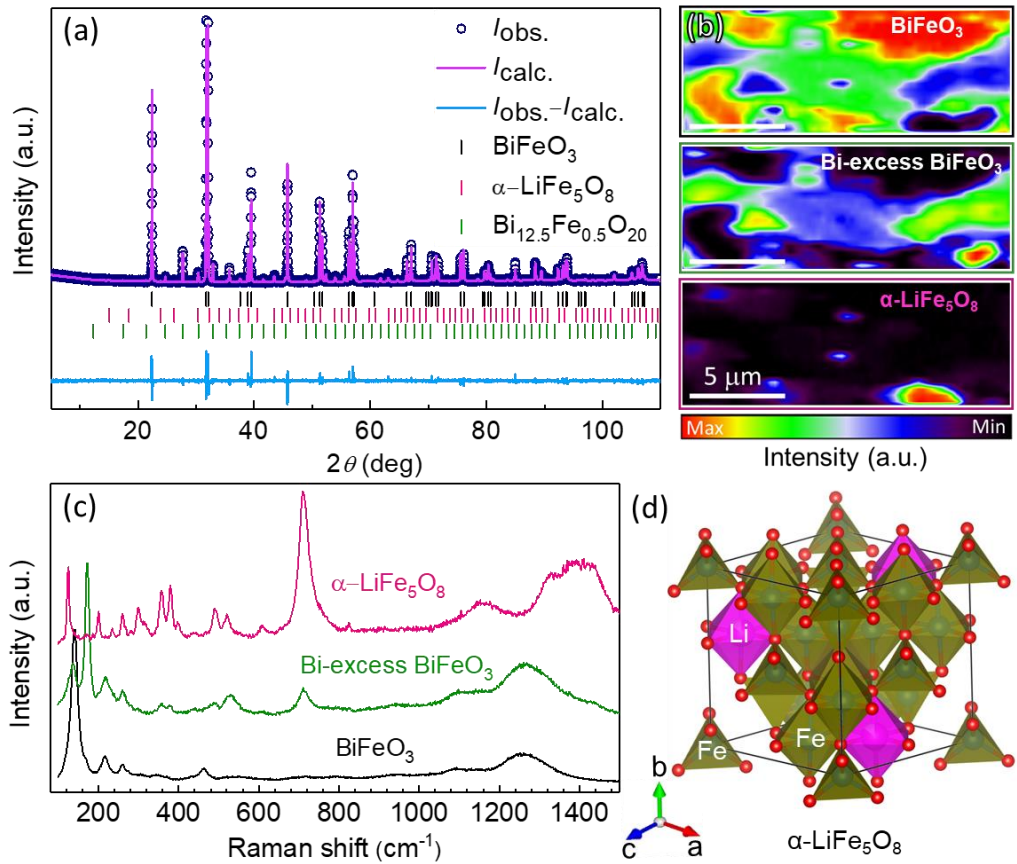

**Figure 1.** Room temperature X-ray powder diffraction and microscopic Raman mapping. (a) Rietveld refinement of powder X-ray diffraction (XRD) data collected from Li$_{0.09}$Bi$_{0.91}$FeO$_3$ (9Li-

BFO) sample. The XRD pattern can be fitted by the three different phases; rhombohedral BiFeO$_3$ (BFO, black), sillenite Bi$_{12.5}$Fe$_{0.5}$O$_{20}$ (green) and ordered spinel α-LiFe$_5$O$_8$ (LFO, pink). (b) Micro-Raman mapping on 9Li-BFO shows the different phase-domains. (c) The Raman mapped area is comprised of three types of spectra from BFO, Bi-excess BFO, and LFO phases. The observation of additional modes at 358, 490, and 530 cm$^{-1}$ of broader linewidth in Raman spectrum of the Bi-excess BFO suggests the presence of the Bi$_{12.5}$Fe$_{0.5}$O$_{20}$ phase. The observed phonon modes of LFO phase are consistent with the Raman spectrum of bulk single crystal of ordered LFO (ref. 40). (d) The crystal structure of ordered spinel LFO (denoted as Fe[Li$^{1+}_{0.5}$ Fe$^{3+}_{1.5}$]O$_4$) shows a specific 1:3 ordering of Li$^{1+}$ and Fe$^{3+}$ at the octahedral *B* sites.

To understand the impact of these coexisting phases, the BFO-LFO ceramic composites are studied by magnetization measurements at various length scales. Figure 2a shows the bulk magnetic properties of the BFO, 3Li-BFO, and 9Li-BFO samples at room temperature. While the undoped BFO appears paramagnetic, the Li doped samples exhibit nearly saturated ferrimagnetic-like M–H loops where the magnetic moment drastically increases with increasing Li concentration. Macroscopically, the difference in magnetic response between the three samples is clearly visible (see supporting movie comparing the samples' response to meet an approaching permanent magnet). To gain insight into the mechanism of these macroscopic responses, local magnetic properties of 3Li-BFO and 9Li-BFO samples are analyzed by magnetic force microscopy (MFM). Nano- to micrometer size magnetic domains are observed in MFM phase images of 3Li-BFO (Figure 2c) and 9Li-BFO (Figure 2e). The density of magnetic domains increases with increasing doping concertation. The line-profile taken along one of the nanodomains further confirms the strong magnetic response from the LFO phase in the nanocomposite ceramic sample (Figure 3f,g). The magnetization measurements suggest that the observed room temperature magnetism in the Li doped samples originate due to the presence of ferrimagnetic spinel LFO phase domains, which is consistent with the X-ray and Raman measurements.

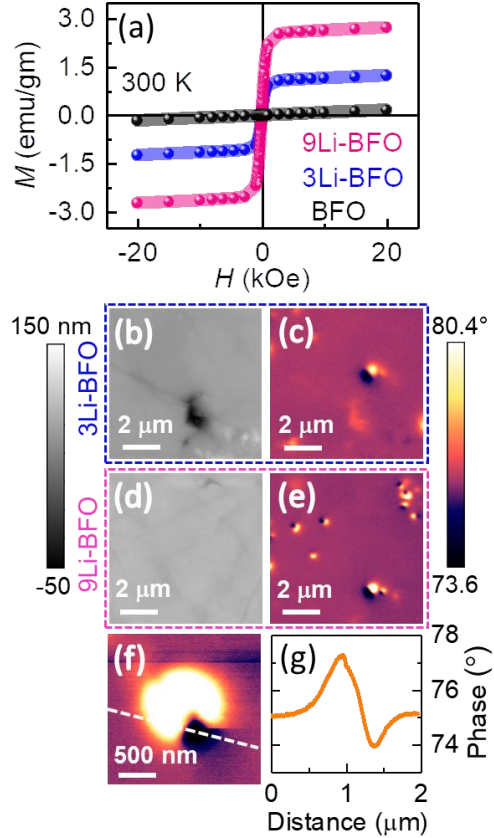

**Figure 2.** Local and bulk magnetic properties. (a) Magnetic hysteresis loops show that the magnetic behavior drastically changes with increasing Li concentration, from nearly paramagnetic undoped BFO to ferri/ferromagnetic Li doped BFO. Scanning probe microscopy of 3Li-BFO (b,c) and 9Li-BFO (d,e) provides surface topography (b,d) and magnetic force microscopy (MFM) phase images (c,e). From MFM images, the nanoscale magnetic domains can be seen embedded in the BFO matrix. The density of magnetic domains increases with increasing Li doping concentration, which significantly enhances the bulk magnetization. These magnetic domains are solely due to the presence of spinel LFO-phase. The magnetic domain size is between 200 nm to 1 µm. (f,g) Line profiles of the MFM phase signal confirms the strong magnetic response from the LFO domains.

Mössbauer spectrometry measurements were performed to gain deeper insight into the local role of Fe valence and phase composition that drive the observed macroscopic magnetic behaviors. Figure 3 shows the Mössbauer spectra collected at 296 and 425 K for the 9Li-BFO sample. The spectra include three main phases—all of which are purely Fe (III) and consistent with X-ray results. Specifically, the material contains a small paramagnetic component (9 atomic % of Fe), a minor magnetic component (16 atomic % of Fe), and a major magnetic component (75 atomic % of Fe). The deconvoluted spectra are consistent with previous reports, which enables the small paramagnetic component to be attributed to $Bi_{12.5}Fe_{0.5}O_{20}$,[42] the minor magnetic component to the ordered spinel LFO phase,[43] and the major magnetic component to antiferromagnetic BFO.[44] The confidence in the presence of the LFO-phase comes from the temperature dependence of the hyperfine magnetic field which collapse more slowly in LFO than in BFO between 296 K to 425 K, thereby leading to the separation of the spectra.

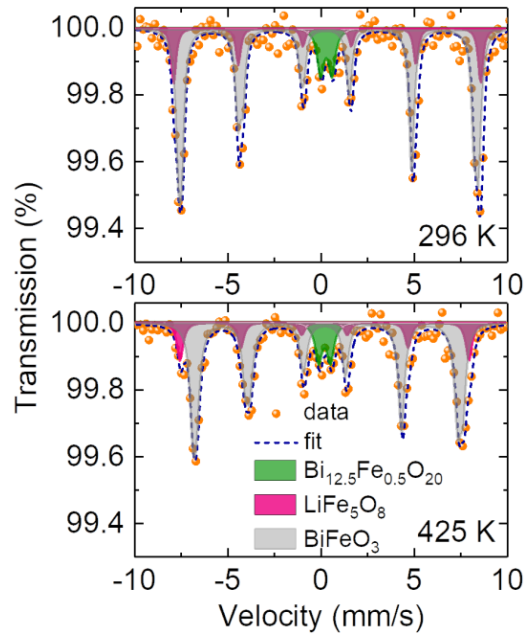

**Figure 3.** Local magnetization and valence state of Fe ions. Mössbauer spectroscopy of the 9Li-BFO sample at (a) 296 K and (b) 425 K temperatures. The fits indicate the presence of ordered spinel LFO, antiferromagnetic BFO, and paramagnetic $Bi_{12.5}Fe_{0.5}O_{20}$ phases. The temperature dependence serves to confirm the presence of spinel LFO as the hyperfine field in LFO collapse more slowly than in BFO between 296 K and 425 K. All three phases exhibit purely trivalent Fe.

Combined band excitation piezoresponse force microscopy (BE-PFM)[45] and MFM measurements were applied on the same location of the 9Li-BFO composite sample to observe coexisting ferroic orders at room temperature. Unlike single frequency PFM, BE-PFM avoids topographic crosstalk by tracking resonance frequency, which rules out the strong dependence of contact resonance frequency on the elastic properties of the different surfaces (phases) in the samples.[46] The results of the overlapping BE-PFM and MFM images are shown in Figure 4a-c. Magnetic domains in the MFM image correspond to the regions in the PFM images showing no electromechanical response (i.e. no amplitude and undefined phase response). These observations show the presence of two independent ferroic phases; magnetically active LFO and ferroelectrically active BFO in our composite sample. Small ferroelectrically inactive regions are also visible and are likely small pockets of the sillenite $Bi_{12.5}FeO_{20}$ phase.

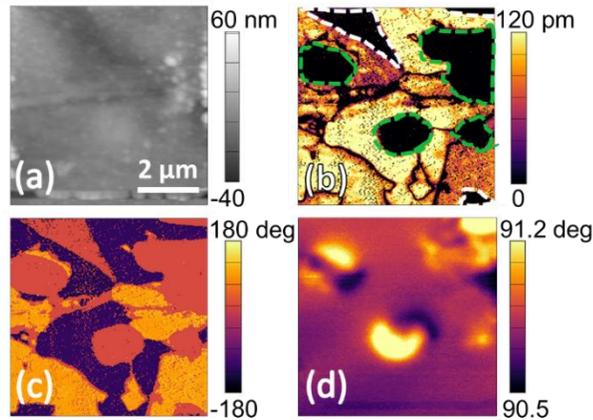

**Figure 4.** Overlapping scanning probe imaging shows coexisting ferroic orders at room temperature. (a) Topography, (b) amplitude and (c) phase images recorded using band excitation piezoresponse force microscopy (BE-PFM). (d) Magnetic force microscopy (MFM) image taken over the same area of the 9Li-BFO sample as BE-PFM images. There is no amplitude response with the undefined phase contrast in those areas (marked by green-dashed) where magnetic domains reside in MFM image. This indicates the presence of two independent ferroic phases in the material—magnetic LFO and ferroelectric BFO. White-dashed indicates ferroelectrically inactive regions that are likely related to pockets of the sillenite $Bi_{12.5}FeO_{20}$ phase.

Time-of-flight secondary ion mass spectroscopy (ToF–SIMS) was used to investigate the local chemical composition and phase separation between LFO and BFO. The chemical sensitivity of SIMS allows the detection of the ion distribution on the surface of the sample with a spatial resolution of ~120 nm.[47,48] Figure 5a demonstrates an averaged mass spectrum with all the base elements $Bi^+$, $Fe^+$, and $Li^+$ present. The distribution of the corresponding peak area as a function of spatial location allows characterization of local chemical changes in the studied area.[49–51] Figure 5b-d shows element specific chemical mappings of Bi$^+$, Fe$^+$, and Li$^+$ ions. The Fe$^+$ map shows that the relative concentration of Fe is higher in the LFO phase domains (Figure 5c). Whereas the Li$^+$ map indicates that the Li resides only in the Fe-rich domains (Figure 5d). The chemical imaging confirms the existence of the Li- and Fe-rich domains of LFO phase randomly distributed in the BFO matrix. The cation overlay (Bi$^+$+Fe$^+$+Li$^+$) image shown in Figure 5e further confirms this phase separation.

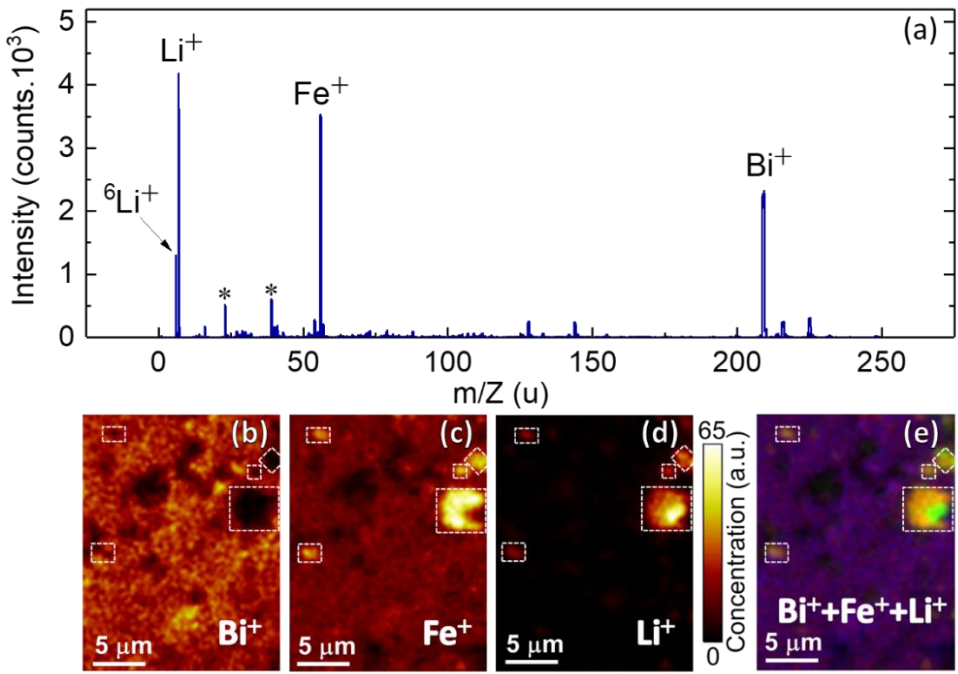

**Figure 5.** Chemical imaging of bulk cermaic nanocomposite sample. (a) Full averaged mass spectrum confirm presence of Bi, Fe, and Li. Note that Na$^+$ and K$^+$ signals (*) are extrinsic surface contamination unrelated to sample composition. (b-d) chemical maps of spatial distribution of base elements Bi, Fe, Li in positive ion detection mode measured on 9Li-BFO sample by ToF-SIMS. The Fe$^+$ map shows that the relative concentration of Fe is higher in the LFO phase domains and the Li$^+$ map indicates that the Li only resides in Fe-rich domains. (e) RGB cation overlay (Fe$^+$ - red, Li$^+$ - green, Bi$^+$ - blue) image confirms the phase separation.

Many multiferroic applications require a form factor other than a bulk ceramic. Using the 9Li-BFO composite ceramic as a target material, it is possible to grow BFO-LFO self-assembled nanocomposite epitaxial thin films using pulsed laser epitaxy (PLE). Films of 42 nm thickness are deposited on single crystal (001) STO substrates. X-ray diffraction measurements indicate the epitaxial growth and the presence of two phases in these films, where the peaks corresponding to both LFO and BFO phases can be seen in $\theta$–$2\theta$ scans of the film (Figure 6a). Epitaxial films with a thickness of 40 nm on (001) STO substrate having 8 nm thick $SrRuO_3$ (SRO) bottom electrodes were also grown by PLE. The AFM topography image shown in Figure 6b indicates a self-assembled square-like nanopillar morphology embedded into the BFO matrix. The local polarization switching and PFM phase and amplitude images of poled domains confirm the film is ferroelectric at room temperature (Figure 6c,d). Similar to the bulk ceramics, no phase or amplitude responses are observed in the nanopillar regions. The MFM and Kelvin probe force microscopy (KPFM) images recorded at room temperature after PFM domain-switching clearly indicate magnetism associated with the LFO-nanopillars (see supporting information). The local magnetic and piezolectric responses confirm that the films are comprised of magnetic LFO nanopillars embedded in a BFO matrix, with a relative volume ratio of 15/85, calculated from the surface topography image (Figure 6b). The small area PFM and MFM images in Figures 6e-h demonstrate that the ferroic properties of spinel LFO and BFO are retained in the heterostructure. Local ferroelectric switching behavior of the BFO-LFO/SRO/STO films is demonstrated with interferometric PFM spectroscopy measurements[52,53] by acquring phase and amplitude hysteresis loops (Figure 6i-j). The PFM phase response shows a complete 180° switching at ±2 V (Figure 6i). The saturation of the PFM amplitude above this voltage (Figure 6j) indicates complete polarization switching in the ferroelectric phase. Room temperature SQUID magnetometry shows that the films' macroscopic behavior is ferromagnetic (Figure 6k). The field-dependent magnetization loops also demonstrate a strong magnetic anisotropy with an in-plane easy axis of magnetization.

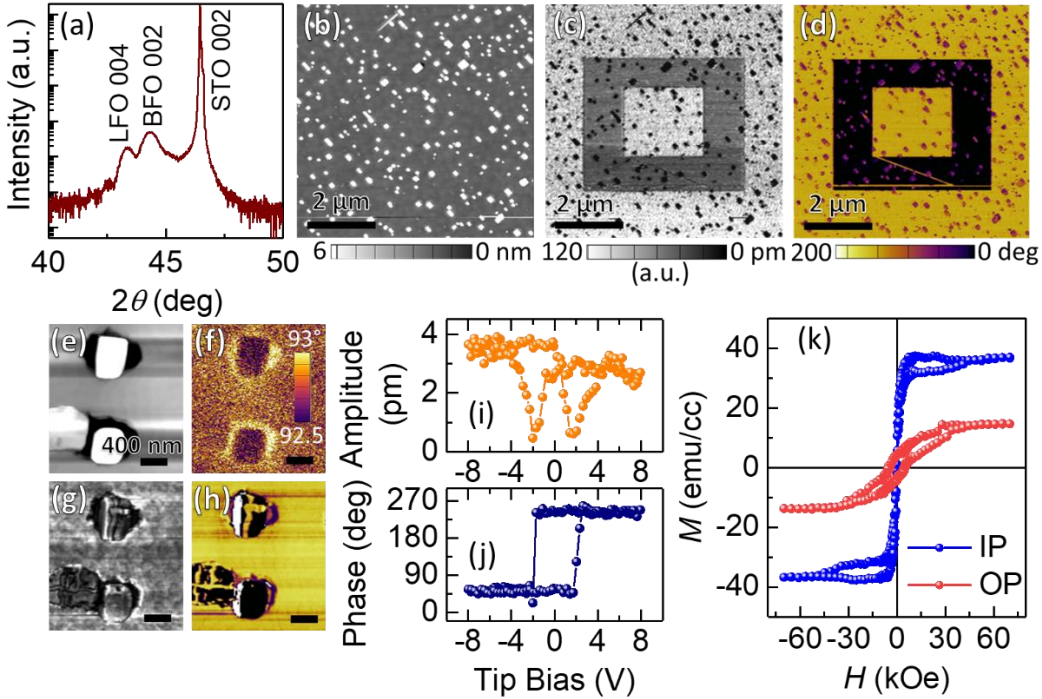

**Figure 6.** Self-assembled nanocomposite thin films of room temperature multiferroic $BiFeO_3$-$LiFe_5O_8$ (BFO-LFO). (a) X-ray diffraction line scan of 42 nm thick BFO-LFO on 001 $SrTiO_3$ (STO, marked by asterisk). The distinct diffraction peaks from BFO and LFO are indexed. (b) AFM topographic image indicates that the LFO nanopillars are embedded within the BFO matrix. (c) Out-of-plane PFM (OP-PFM) amplitude and (d) phase images taken after poling the nanocomposite film using ± 6 V. The PFM measurements were performed on a 40 nm thick BFO-LFO film grown on 8 nm $SrRuO_3$ coated STO. Like bulk ceramics, the phase and amplitude responses are suppressed in LFO nanopillar areas. (e-h) Enlarged view of the local magnetic and ferroelectric response of BFO-LFO films grown directly on STO; the (f) MFM and (g,h) PFM images acquired over the same area of the sample show that the nanopillars have a strong magnetic response but are not ferroelectric. The local ferroelectric switching is confirmed by PFM (i) amplitude and (j) phase hysteresis loops. (k) The in-plane (IP) and out-of-plane (OP) magnetic hysteresis loops measured at 300 K indicate the anisotropy of magnetization in the BFO-LFO nanocomposites film.

3D ToF-SIMS is used to examine the local chemical composition of the BFO-LFO/STO films. In these measurements, a $Bi^{3+}$ primary ion beam is combined with a $Cs^+$ sputter beam to access element specific depth profiling. Figures 7a and b show that Li is locally confined to nanopillars of LFO phase and not dilutely distributed in the BFO matrix. A strontium ion ($Sr^+$) concentration map is generated to locate the substrate (Figure 6c). Overlaying the Sr and Li maps shows a depth profile image in which the LFO nanopillars are normal to the substrate (Figure 6d).

Combined with the scanning probe work, these measurements provide a full structural and functional map of the self-assembled vertically aligned nanocomposite BFO-LFO films.

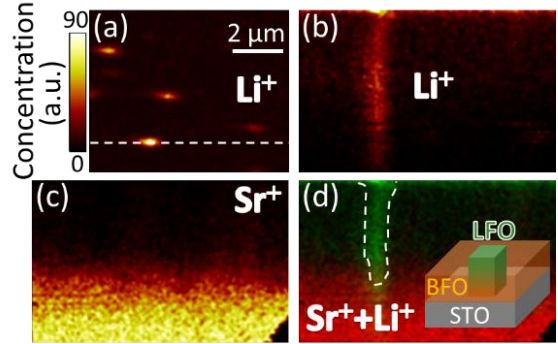

**Figure 7.** 3D chemical imaging of self-assembled vertically aligned nanocomposite film. ToF-SIMS compositional mapping of BFO-LFO/STO film. (a) Lateral (X-Y) surface view shows pockets of Li rich material. (b) Vertical (X−Z) cross-sectional depth view taken along the white-dashed line in (a). The overlay image of Sr$^+$ (red) from the STO substrate and Li$^+$ (green) ions is shown in (d), which confirms that the LFO nanopillars are vertically aligned within the BFO matrix.

To gain insight into the mechanism of how these phases form and coexist, the energetics of phase decomposition and formation stability of the LFO phase as a function of Li doping were modeled using first principles density functional theory (*DFT*). The formation energies (Δ*E*) are computed for BFO, LFO, and for Li doped BFO, where Li sits on both Fe (Li$_{Fe}$) and Bi (Li$_{Bi}$) sites as well as interstitially (Li-int), as follows:

$\Delta E_{BFO} = E_{BFO} - \frac{1}{2} E_{Bi_2O_3} - \frac{1}{2} E_{Fe_2O_3}$

$\Delta E_{LFO} = E_{LFO} - \frac{1}{2} E_{Li_2O} - 5/2\ E_{Fe_2O_3}$

$\Delta E_{Li_{Bi}\text{-}BFO} = E_{Li_{Bi}\text{-}8BFO} - 4\ E_{Bi_2O_3} - 7/2\ E_{Fe_2O_3} - \frac{1}{2} E_{Li_2O} - \frac{1}{2} O_2$

$\Delta E_{Li_{Fe}\text{-}BFO} = E_{Li_{Fe}\text{-}8BFO} - 7/2\ E_{Bi_2O_3} - 4\ E_{Fe_2O_3} - \frac{1}{2} E_{Li_2O} - \frac{1}{2} O_2$

$\Delta E_{Li\text{-int -}BFO} = E_{Li\text{-int-}8BFO} + \frac{1}{2} O_2 - 4\ E_{Bi_2O_3} - 4\ E_{Fe_2O_3} - \frac{1}{2} E_{Li_2O}$

where Δ*E* are the total energies computed from *DFT* of the individual species. All formation energies are computed relative to the binary oxides: Bi$_2$O$_3$, Fe$_2$O$_3$, and Li$_2$O, where Li substitutional and Li interstitial systems are considered in a 2 x 2 x 2 supercell of BFO. G-type, A-type, and C-type antiferromagnetic orderings and ferromagnetic and nonmagnetic states were considered for BFO. As in previous doping studies, only G-type ferromagnetic orderings were considered for the doped-BFO systems.[54] Ferromagnetic and nonmagnetic orderings as well as multiple ferrimagnetic configurations were considered for LFO.[55] The relative energetics and ground state lattice constants for the compounds were calculated using *DFT* (see Table S3-S6, supporting information). In the final analysis, only the lowest energy magnetic configurations were considered.

Table 1 lists the formation energies for BFO, LFO, Li substituted BFO (on both the Bi and Fe site), and Li interstitials in BFO relative to the bulk binary oxides. LFO is the most stable phase of the BFO-based compounds while Li prefers to be incorporated as an interstitial rather than to substitutionally dope either Bi or Fe sites in BFO.[54] To better understand the tendency to form LFO over BFO, the ternary phase diagram for the formation of LFO, BFO, and Li-interstitial BFO was examine as a function of the concentration of binary oxides, $Li_2O$, $Bi_2O_3$, and $Fe_2O_3$. Here, Li-substituted compounds are omitted, due to Li's preference to be incorporated as an interstitial.

**Table 1** The formation energies for Li substituted BFO (on both the Bi and Fe site), Li interstitials in BFO, BFO and LFO.

| | Formation Energy eV/Fe (or *B*-site) |
|---|---|
| $\Delta E_{Li_{Bi}\text{-BFO}}$ | -0.263 |
| $\Delta E_{Li_{Fe}\text{-BFO}}$ | -0.301 |
| $\Delta E_{Li\text{-int-BFO}}$ | -0.338 |
| $\Delta E_{BFO}$ | -0.223 |
| $\Delta E_{LFO}$ | -0.354 |

The phase stability maps for the formation of LFO, BFO, and Li-interstitial BFO are shown in Figure 8a-c. To allow for an equal comparison, these energies are normalized by the fraction of the binary oxides in a region of the ternary phase diagram. As expected, LFO is stable in regions with small amounts of $Bi_2O_3$ and large amounts of $Fe_2O_3$, while BFO and Li-interstitials are more stable near regions with roughly 0.5 $Bi_2O_3$ and minimal $Fe_2O_3$ and $Li_2O$. Importantly, the formation of LFO from the end member binary oxides is substantially more stable than either BFO or Li-interstitial BFO, which is consistent with the experimental results. To assess the energetics of phase composition, an overview map of different phases as a function of $Bi_2O_3$, $Li_2O$, and $Fe_2O_3$ is presented as Figure 8d. These results show that under most conditions it would be expected that LFO will form. However, there is a small region at low $Li_2O$ concentrations where it can be expected that Li-interstitials may arise in bulk BFO.

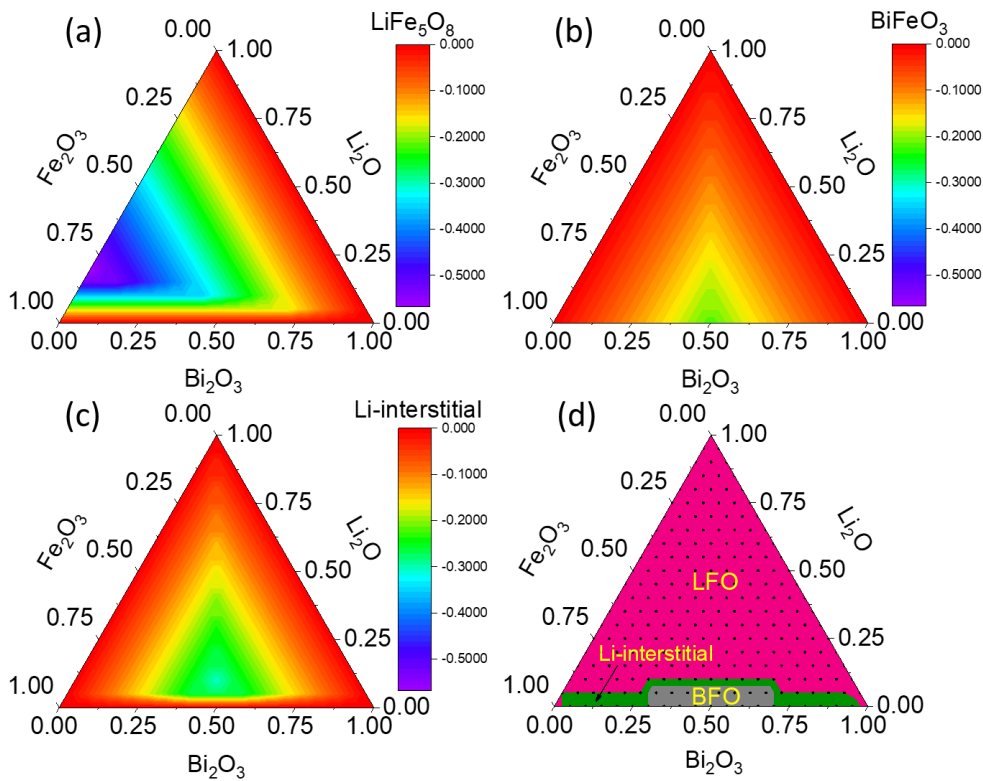

**Figure 8.** Energetics of phase decomposition due to Li doping in BFO. (a-c) The phase stability maps for the formation of LFO, BFO, and Li-interstitial BFO. (d) An overview map of different phases as a function of $Bi_2O_3$, $Li_2O$, and $Fe_2O_3$. The formation of LFO from the end member binary oxides is substantially more stable than either BFO or Li-interstitial BFO, which is consistent with the experimental results.

## CONCLUSIONS

This work introduces a new approach to create a robust room temperature multiferroic. Lithium doping in $BiFeO_3$ (BFO) favors the formation of the ferrimagnetic B-site-ordered α-$LiFe_5O_8$ (LFO) spinel as a secondary phase during the synthesis of $Li_xBi_{1-x}FeO_3$ ceramics. The energetics of phase formation and phase separation are supported by first principle calculation, showing that formation of the LFO phase is energetically favorable for Li doping into BFO. The multiferroic properties of bulk ceramics and thin films of BFO-LFO nanocomposites are examined using multimodal imaging methods at various length-scales. Self-assembled epitaxial BFO-LFO nanocomposite films are found to be comprised of LFO nanopillars embedded within a single crystal BFO matrix. Strong magnetism and ferroelectricity are observed at room temperature. This work provides an approach for the design of self-assembled vertical heteroepitaxial nanostructures that may find use in future energy, sensing, and memory applications.

## METHODS

**Bulk ceramic preparation and film growth:** Phase pure $BiFeO_3$ (BFO) and $Li_xBi_{1-x}FeO_3$ (x = 0.03; 3Li-BFO, and x = 0.09; 9Li-BFO) ceramic targets were synthesized by a conventional solid-state reaction method using $Bi_2O_3$, $Fe_2O_3$ and $Li_2O$ powder precursors (Alfa-Aesar >99.99%). The powder precursors were weighed in stoichiometric amounts according to different target compositions followed by mixing in ball milling for up to 4 h. The powder mixtures were calcined at furnace temperature of 760 °C in air for 1.5 h. After calcination, these powders were pressed into pellets using a cold isostatic method (~32 MPa) and sintered at 780 °C in air for 1 h to form polycrystalline ceramic targets. Using 9Li-BFO as ceramic target, thin films of BFO-LFO nanocomposites were deposited on single-crystal SrTiO$_3$ (001) substrates by pulsed laser epitaxy. The KrF excimer laser (248 nm in wavelength) at a laser fluence of 1.5 Jcm$^{-2}$ and at a laser repetition rate of 5 Hz was used to ablate the ceramic target. During deposition, the substrate temperature and oxygen partial pressure were 660°C and 80 mTorr, respectively. After deposition, the samples were cooled to room temperature at higher oxygen pressure, ~100 Torr.

**Characterization of Structure and Properties:** Powder x-ray diffraction data were collected in a PANalytical X'pert Pro MPD diffractometer with Cu $K_{\alpha 1}$ radiation (wavelength $\lambda$ of 1.540596 Å). The full-pattern Rietveld refinements were carried out for the diffraction data within the program *Fullprof*.[56] The phase contents were also quantitatively estimated from their scale factors, which were obtained from the Rietveld refinements.[57,58] The structural characteristics of BFO-LFO (001) thin films were analyzed by a four-circle, high-resolution X-ray diffractometer (X'Pert Pro, PANalytical) using Cu $K_{\alpha 1}$ radiation. Micro-Raman measurements were performed in a Renishaw 1000 confocal Raman microscope using a 532 nm diode-pumped solid-state laser (Cobolt) with a laser power 10 mW. Raman spectra were collected in back scattering geometry using 100× objective (NA= 0.75). For Raman mapping, we collected 2348 spectra (10s integration time, and using 500 nm step size) on a 15×7 µm$^2$ region of the sample. Mössbauer spectra of 9Li-BFO of a 24 mg/cm$^2$ powder sample were acquired using the $^{57}$Fe Mössbauer resonance, employing a Wissel constant acceleration drive calibrated with α-iron at 296 and 425 K with the sample placed in a Janis SHI-850 closed-cycle cryo-oven. Macroscopic magnetic hysteresis loops were performed by using a magnetic property measurement system (Quantum Design). SPM studies were performed with a commercial AFM system (Asylum Research Cypher) equipped with an integrated quantitative Laser Doppler Vibrometer (LDV) system (Polytec GmbH, Waldbronn, Germany). MFM images were obtained with a magnetic tip working at lift mode with a lift height of 50 nm (for films) and 150 nm (for ceramics) using a permalloy-coated Si-tip (ASYMFMLC-R2, Asylum research). PFM measurements were carried out using a conducting PtSi-coated tip (PtSi-FM-20, Nanosensors). BE-PFM was implemented using NI-5122 and NI-5412 cards (National Instruments) controlled by custom Matlab software (MathWorks). ToF-SIMS measurements were performed using TOF.SIMS.5-NSC (IONTOF GmbH, Germany) instrument. We used a $Bi_3^+$ liquid metal ion gun (energy 30 keV, current 0.5 nA, spot size ~120 nm) as a primary ion source and Cs+ ion gun as a sputter source (energy 1 keV, current 75 nA, spot size ~20 μm). Secondary ions were analyzed using time-of-flight mass analyzer operated in positive ion detection mode with mass resolution $m/\Delta m = 100 - 300$ to track peaks of $Fe^+$, $Li^+$, $Bi^+$ and $Sr^+$. Surface measurements were carried out using $Bi_3^+$ source only, rastered over 20 x 20 μm area with 256 x 256 px resolution. 3D measurements were performed in non-interlaced mode, when each scan by $Bi^{3+}$ primary source (20 x 20 μm area) was followed by sputtering using $Cs^+$ source for 2 s (over 100 x 100 μm area). Acquired three-dimensional maps of the peaks' spatial distribution were used to identify local changes in the chemistry of the studied sample.

*DFT* **Calculations:** The first principles *DFT* calculations were performed with the local density approximation (LDA) for exchange-correlation as implemented in the Vienna *Ab initio* Simulation Package (VASP).[59–61] For all calculations we found a 500 eV cutoff was sufficient to ensure converged results. We employed projected augmented wave (PAW) potentials with the following electronic configurations: Bi $5d^{10}6s^26p^3$, Fe $3p^63d^64s^2$, Li $1s^22s^1$, O $2s^22p^4$.[62] For each species, we optimized both the internal atomic coordinates and the lattice constants until the forces and external pressure on the cell were less than 0.01 eV/Å (on each atom) and 1 kB, respectively. For Fe containing compounds, we employed a simplified rotationally invariant Hubbard parameter $U$ = 5 eV to account for the strong correlations.[63] This value was chosen as it provided the best balance for comparison between the $LiFe_5O_8$ and $BiFeO_3$ electronic structures. Similar values were reported to adequately account for the weak magnetism in BFO.[64] For each system we used Monkhorst-Pack k-point meshes as listed in Table S3 (supporting information). We note that the reference chemical potential for oxygen $\mu(O)$ was determined from the atomization energy of $O_2$ employing a spin polarized calculation in the spin 3/2 state: 1/2 $E_{total}(O_2)$ = -5.23 eV. This is in good agreement with the experimental $\mu(O)$ = -5.17 eV.[65] We do not apply a correction to $\mu(O)$, as in the previous study on Li-doped BFO.[54] For all Li substitutions we only consider neutral Li dopants and G-type AFM ordering for BFO. While Li-interstitials were determined to be the most stable Li-doping scenario, we observe that our computed relative energetics for Li-substituted BFO unit cells, reproduce previous GGA calculations which indicate preferential Li substitution onto the Fe-site rather than Bi-sites.[54] We note a difference in the magnitude of the formation energy of 0.11 eV[54] versus 0.30 eV per 2 x 2 x 2 BFO unit cell. This number is within the estimated error for PBE predictions of formation energies.[66]

## ASSOCIATED CONTENT

**Supporting Information.** Results of the X-ray diffraction, MFM and KPFM images after electrical poling, and details of DFT+U calculations. This material is available free of charge via the Internet at http://pubs.acs.org.

## AUTHOR INFORMATION

**Corresponding author**
sharmay@ornl.gov and wardtz@ornl.gov
Notes: The authors declare no competing financial interest.

**Corresponding author**
sharmay@ornl.gov and wardtz@ornl.gov
Notes: The authors declare no competing financial interest.

ACKNOWLEDGMENTS

This work was supported by the US Department of Energy (DOE), Office of Science, Basic Energy Sciences (BES), Materials Sciences and Engineering Division. Scanning probe microscopy and Raman spectroscopy studies were performed as user projects at the Center for Nanophase Materials Sciences, which is sponsored at Oak Ridge National Laboratory (ORNL) by the Scientific User Facilities Division, BES, DOE. QZ acknowledges the support of the Center for Emergent Materials, an NSF MRSEC, under Award Number DMR-1420451. The first principles DFT calculations were performed at the National Energy Research Scientific Computing Center, a DOE Office of Science User Facility supported by the Office of Science of the U.S. Department of Energy under Contract No. DE-AC02-05CH11231.

**Table of content graphic:**

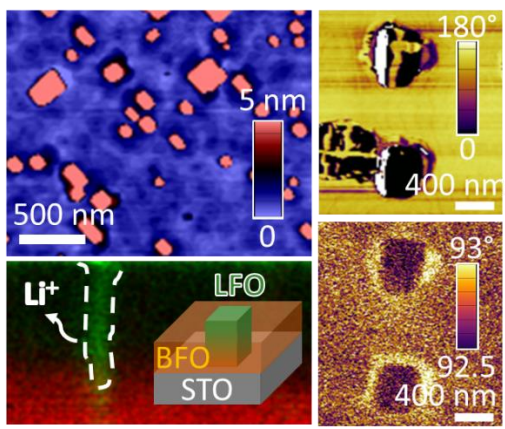

Room temperature multiferroic self-assembled vertically aligned BiFeO$_3$-LiFe$_5$O$_8$ nanocomposite film.